\documentstyle[12pt]{article}

\def\be {\begin{eqnarray}}
\def\ee {\end{eqnarray}}
\def\ni {\noindent}
\def\vpar {v_\parallel}
\def\vperp {v_\perp }
\def\vk {V_{\bf K}}
\begin{document}
\setcounter{page}{1}
\hfill {NORDITA-93/72 A/S/N}
\vskip 0.4in
\begin{center}
{\large\bf Neutrino Pair Bremsstrahlung in \\
Neutron Star Crusts: a Reappraisal }
\end{center}
\vskip 0.6in
\centerline{C. J. Pethick $^{1,2}$ and Vesteinn Thorsson$^{1}$ \\}
\vskip 0.4in
\centerline{\it $^{1}$ NORDITA, Blegdamsvej 17, DK-2100 Copenhagen \O, Denmark
}
\centerline{\it $^{2}$ Department of Physics, University of Illinois at
Urbana-Champaign,}
\centerline{\it 1110 West Green St., Urbana, Illinois 61801-3080, USA}
\vskip .6in

\centerline{\bf Abstract}
\vskip .5cm
\noindent
We demonstrate that band-structure effects suppress bremsstrahlung of neutrino
pairs by electrons in the crusts of neutron stars at temperatures of the order
of $5\times 10^9 \,{\rm K}$ and below.  Taking this into account, together with
the fact that recent work indicates that the masses of neutron star crusts are
considerably smaller than previously estimated, we find neutrino pair
bremsstrahlung to be much less important for the thermal evolution of neutron
stars than earlier
calculations suggested.
\par\vfill

\newpage

Observations of neutron star temperatures have potential for probing the
interiors of neutron stars, and  in recent years much effort has been devoted
to making such measurements, especially with the Einstein Observatory and
ROSAT.  On the theoretical front there have also been developments regarding
neutrino emission in matter in the cores of neutron stars, which is the main
cooling mechanism early in the life of a neutron star.  In this Letter we
consider bremsstrahlung of neutrino
pairs by electrons scattering from the Coulomb field of
nuclei in the crust of a neutron star,
$e^- + Z  \rightarrow e^- + Z + \nu + \bar{\nu}$.  According to the usual
picture of neutron star cooling, this process can
play an
important role, especially if neutrons are superfluid and/or protons
superconducting\cite{st}. In the latter case,  neutrino emission in
the core
will be suppressed at temperatures below the transition temperatures, and
neutrino pair bremsstrahlung in the crust can dominate.  Even if nucleons in
the interior are normal, the crust bremsstrahlung process has been estimated to
be comparable in importance to the modified Urca process, which for the past
quarter of a century has
been regarded as the ``standard" process.

The theory of the bremsstrahlung process in dense matter was developed by Festa
and Ruderman\cite{fesrud}, and subsequently extended by other workers
\cite{flowers,dicusetal,soybro,itoh}. The basic assumption common to these
treatments is that the electron-ion interaction may be treated in first-order
perturbation theory, and the conclusion is that, for a perfect lattice, the
rate of emission of energy by neutrinos varies approximately as $T^6$. What we
demonstrate in this Letter is that the process is suppressed exponentially at
low temperatures.  The physical reason for this is that repeated interactions
of an electron with the lattice give rise to splittings between bands which can
range up to $1 \, {\rm MeV}$.  Consequently, at temperatures of order
$10^9 \,{\rm K }\sim 0.1 \,
{\rm MeV}$,  at which the neutrino bremsstrahlung process has been thought to
be
important, the rate of the process is much less than previously estimated. We
begin by describing the microscopic calculations of the bremsstrahlung rate,
and then, incorporating recent developments in the theory of matter in the
crust of a neutron star, we explore implications for the thermal evolution of
neutron stars.

To set the scene, let us examine the electron spectrum.  In most of the
crust of a neutron star, electrons move in a periodic lattice
of nuclei.  At the lower densities, the nuclei are essentially spherical, and
the lattice bcc, but at densities approaching that at the inner boundary of
the crust, nuclei may be rod-like or plate-like.  For densities above $\sim
10^6 \,{\rm g}\,{\rm cm}^{-3}$ the electrons are relativistic.  As we shall
demonstrate, splittings between bands are generally small compared with
electron energies, and therefore they may be estimated in the
nearly-free-electron approximation.  Since the Fermi energy is much greater
than the
electron rest mass, it is an excellent approximation to work in the extreme
relativistic limit, in which the electron helicity is a good quantum number.
This simplifies significantly the calculations, and the errors introduced are
small, of order $(m_e c^2/\mu_e)^2$.

The crystal potential has most effect on states for which the
free particle energies $\epsilon_{\bf p}$ and $\epsilon_{{\bf p}-\hbar{\bf K}}$
are almost equal for some reciprocal lattice vector $\bf K$.
The energy eigenvalues for the upper and lower bands, denoted by $+$ and $-$
respectively, are given by
\be
E^\pm({\bf p})
&=& \frac{
\epsilon_{\bf p} + \epsilon_{ {\bf p}-\hbar{\bf K} }
}{2}
\pm \sqrt{
(\frac{
\epsilon_{\bf p} - \epsilon_{ {\bf p}-\hbar{\bf  K} }
}{2})^2
+ |\vk|^2 } \,\, \,\,,
\label{eenergy}
\ee

\ni and the corresponding states are

\be
\Psi^+_{{\bf p},\sigma}({\bf r}) &=&
u_{\bf p} e^{\frac{i}{\hbar} {\bf p}\cdot{\bf r}} u_\sigma ({\bf p})
+ v_{\bf p} e^{\frac{i}{\hbar} ({\bf p}-\hbar {\bf K})\cdot{\bf r}}
u_\sigma ({\bf p}-\hbar {\bf K}) \nonumber\\
\Psi^-_{{\bf p},\sigma}({\bf r}) &=&
v_{\bf p} e^{\frac{i}{\hbar} {\bf p}\cdot{\bf r}} u_\sigma ({\bf p})
- u_{\bf p} e^{\frac{i}{\hbar} ({\bf p}-\hbar {\bf K})\cdot{\bf r}}
u_\sigma ({\bf p}-\hbar {\bf K})\,\,,
\label{estates}
\ee

\ni where $u_\sigma$ is a ( four-component ) spinor of helicity $\sigma$,
and the ``coherence factors" are given by

\be
u_{\bf p}^2 = \frac12 (1+\frac{ \xi_{\bf p} }{{\cal E}_{\bf p}})\,\,,\,\,
          v_{\bf p}^2 = \frac12 (1-\frac{ \xi_{\bf p} }{ {\cal E}_{\bf p} }) \,
       \,\,  ,\,\,{\rm and} \, u_{\bf p} v_{\bf p} = \frac{\vk}{2 {\cal E}_{\bf
p}}\,\,\,\,,
\ee
\ni with  $\xi_{\bf p} =
( \epsilon_{\bf p} - \epsilon_{ {\bf p}-\hbar{\bf K} })/2$ and
${\cal E}_{\bf p} = \sqrt{ \xi_{\bf p}^2 + |\vk|^2 }$.

These results are essentially the same as for the non-relativistic case, except
that the free-particle dispersion relation is that for massless particles, and
the matrix element of the electron-lattice interaction is modified.
The splitting between bands due to the periodic potential is
$2 |\vk|$, where

\be
\vk =
- \vperp \frac{4 \pi e \rho_{\bf K}}{K^2} \,\,\,\,\,.
\ee

\ni In this equation $v_{\perp} = \sqrt{1 - (K/(2k_F))^2}$,
with $k_F \simeq \mu_e/(\hbar c)$ being the Fermi wavenumber. The quantity
$\rho_{\bf k}$ is the Fourier transform of the total charge density,
$ \rho_{\bf k} =  e n_p F({\bf k}) / \varepsilon ({\bf k})$, and $n_p$
is the proton density. Here $F({\bf k})$ is the form factor, which has
contributions from the shape of the nuclear charge distribution as well as
from thermal vibrations (the Debye-Waller factor), and
$\varepsilon = 1 + k_{FT}^2/k^2$ is the static dielectric function,
where $k_{FT} = \sqrt{4 \alpha /\pi } \, k_F $
is the Fermi-Thomas screening wavenumber,
with $\alpha = e^2/(\hbar c)$. The
factor $v_{\perp}$ is essentially the overlap
between helicity states familiar in calculations of scattering
of relativistic electrons.
For point-like nuclei, with atomic number $Z$, and for the smallest reciprocal
lattice vector in a bcc lattice, the splitting is $\simeq 0.018 (Z/60)^{2/3}
\mu_e$.  In the inner crust of neutron stars, electron Fermi energies range
up to $\sim$ 75 MeV, and $Z$ can be as large as $\sim 60$\cite{lrp} so
splittings can be 1 MeV or more.

    Let us now estimate the rate of energy emission.  The basic process is
shown in Fig.1(a).  Here an electron moving in the lattice potential emits a
neutrino-antineutrino pair.  It follows directly from Fermi's golden rule that
the rate of energy emission in neutrino pairs, per unit volume, is given by

\be
\dot{E} =
\frac{2 \pi}{\hbar}
\sum \delta( E_f - E_i )
f_1 ( 1 - f_2 )  | H_{fi} |^2 ( E_{\nu} + E_{\bar \nu} )    \,\,\,\,.
\label{goldrule}
\ee

\ni
Here $f$ is the Fermi distribution function, and the factor $f_1 (1-f_2)$
ensures that
the initial electron state, 1, is occupied, and the final state, 2, vacant.
We assume
that neutrinos can escape freely from matter, and therefore there are no
blocking factors for neutrinos.  The sum is over momenta and
helicities of incoming and outgoing particles.

We shall focus on temperatures low enough that the neutrino momentum is small
compared with any reciprocal lattice vector.  Under these circumstances it is
easy to see that the important processes will be ones involving electrons lying
close to the Fermi surface, and with crystal momenta having a component close
to $K/2$ in the direction of some reciprocal lattice vector $\bf K$.
The electron spectra for
two values of $p^\perp$, the component of $\bf p$ perpendicular to $\bf K$,
are shown
in Fig.2, as a function of $p^\parallel$, the component of ${\bf p}\pm{\bf
K}/2$ parallel to $\bf K$.
If the initial and final electron states are in the same band, the
bremsstrahlung process is kinematically forbidden, as we now show.
On the one hand, the electron (group) velocity $\vec \nabla_p
E$ cannot exceed $c$, and therefore the energy difference, $E_1 - E_2$, between
electron states
is less than $cq$, where $q$ is the total momentum of the neutrino pair.  On
the
other hand, for the neutrino pair, the energy difference must exceed $cq$.
Consequently it is impossible simultaneously to conserve energy and momentum.
For states in different bands there will generally be a finite energy
difference even for small momentum transfers, and so the process is
kinematically allowed.

We now examine the process in which an electron in the
upper band makes a transition to the lower one.  In evaluating matrix elements
of the weak interaction Lagrangian,

\be
{\cal L} = -\sqrt{2} G {\bar \Psi}_\nu
\gamma_\alpha P_L
\Psi_\nu
{\bar \Psi} ( C_L
\gamma^\alpha P_L
+ C_R
\gamma^\alpha P_R )
\Psi \,\,\,\,\,,
\label{weakint}
\ee
\ni
one must use Bloch
electron states Eq.(2), rather than plane waves.
In Eq.(\ref{weakint}), $G$ is the Fermi coupling constant,
$P_{L,R} = (1 \mp \gamma_5 )/2$,
and in terms of the weak mixing angle $\theta_W$,
$C_L = 1 + 2\sin^2 \theta_W$ and $C_R = 2\sin^2 \theta_W $
for the emission of electron neutrinos.
For the emission of
muon and $\tau$ neutrinos,
the corresponding couplings are
$C_L'= - 1 + 2\sin^2 \theta_W$ and $C_R'= 2\sin^2 \theta_W $.
For emission of electron neutrinos one finds

\be
\dot{E} =
\frac{ G^2 }{ 24 \pi^6 \hbar^{10} c }
\frac{ C_A^2 + C_V^2 }{2}
\sum_{\bf K}
\vpar^2 \,
\int dp^\parallel_1 \,\, d^3q \,\,
\omega^2  \,  q^2 \, u_1 v_2 v_1 u_2 \,
\frac{1}{e^{\beta \omega} - 1} \,
\theta( \omega - c |{\bf q}| )    \,\,,
\label{edotgen}
\ee
\ni where $C_V=(C_L+C_R)/2$, $C_A=(C_L-C_R)/2$,
$\vpar = K/(2k_F)$,
$\beta=1/k_B T$, $k_B$ is the Boltzmann constant and $ \omega = E_1 - E_2 $.
The total emission rate from all types of neutrinos is obtained
by replacing
$ C_V^2 + C_A^2$ by
$C_A^2 + C_V^2 + 2 ( (1-C_A)^2 + (1-C_V)^2 ) $.
The result (\ref{edotgen}) is valid for arbitrary values of $T/|{\vk}|$
and to lowest order in $k_B T/\mu_e$.
We remark that processes in which an electron initially in the ``lower" ($-$)
band makes a transition to a state in the ``upper" ($+$)
band is kinematically
forbidden, even though the energy of the initial state can be higher
than that of the final state.

Simple analytical results may be obtained in limiting cases, and we first
consider temperatures small compared with $\vk$, where our results differ
dramatically from earlier ones.
In the low-temperature limit, one may expand the integrand in
Eq.(\ref{edotgen}) in powers of $\xi/\vk$, and one finds

\be
{\dot E}_<
 = \frac{2 G^2}{ 3 \pi^\frac92 \hbar^{10} c^9 }
\frac{ C_A^2 + C_V^2 }{2}
\mu_e
(k_BT)^\frac72 \,\,
\sum_{\bf K}
\frac{(1-\vperp)^{\frac 12} }{
\vperp^{\frac 52}  ( 1 + \vperp)   }
|\vk|^\frac92
e^{-\frac{1}{k_BT} \frac{2|\vk|}{1+v\perp} }  \,,\,  k_B T \ll |\vk|  \,\,\,.
\label{lowtemp}
\ee

\ni
The exponential dependence reflects the fact that the minimum energy of
the neutrino pair is $2|\vk|/(1+v_{\perp})$.  This is easily seen by observing
that the energy of the neutrino and antineutrino is given by $\omega = E_1 -
E_2 \ge 2|\vk| + q_{\perp}v_{\perp}$, and that, in addition, the four-momentum
of the neutrino pair must be time-like, $\omega \ge cq \ge c|q_{\perp}|$.

Next we consider the limit of temperatures high compared with $|\vk|$.
The coherence factors may be expanded in powers of $\vk/\xi$
and the energy emission rate is
\be
{\dot E}_> =
\frac{4 \pi G^2 }{567 \hbar^{10} c^9 }
\frac{C_A^2+C_V^2}{2} \mu_e (k_B T)^6
\,\,\sum_{\bf K} \frac{\vpar}{\vperp^2} ( 1 -
\frac{\vpar^2}{\vperp^2} \log{ \frac{1}{\vpar^2} } ) |\vk|^2
\,,\, k_B T \gg |\vk| \,\,\,\,.
\label{hightemp}
\ee
\ni
This result is consistent with what was found in earlier calculations
\cite{flowers,itoh} in which the electron-lattice interaction was treated
perturbatively.  (See Fig.1(b).)
To exhibit clearly scaling properties, we
neglect all
but the smallest reciprocal lattice vectors, put form factors equal to unity,
and neglect the term in parentheses in Eq.(\ref{hightemp}).  We then find an
emissivity per unit mass from all species of neutrinos of $0.23 \, x \, Z \,
T_8^6 \, {\rm erg}\, {\rm g}^{-1} \, {\rm s}^{-1}$ for spherical nuclei.  Here
$x = n_p/n$ is the proton fraction, where $n$ is the density of nucleons.  Our
results show that the neutrino emissivity due to the component of the lattice
potential with wave-vector $\bf K$ is reduced at low temperatures by a factor
$\sim |\vk/{k_BT}|^\frac52 \exp(-2|\vk|/[k_B T (1+\vperp)])$
compared with the high-temperature expression.

In the crusts of neutron stars many reciprocal lattice vectors contribute to
neutrino emission, because the number of reciprocal lattice vectors for which
$K < 2 k_F$ is $\simeq 4Z$.  At the highest temperatures neutrino emission
will be dominated by the smallest reciprocal lattice vectors, since they
have the largest periodic lattice potential. However their contributions
will be the first to be suppressed by band structure effects as the temperature
is
lowered, and consequently the most important reciprocal lattice vectors for
neutrino emission will increase with decreasing temperature.  To give a sense
of this effect we have constructed a simple interpolation formula for the
neutrino emission which, for each reciprocal lattice vector, agrees with
the exact results in the high- and low-temperature limits.  This is the sum of
Eq.(\ref{hightemp}) with an additional factor
$\exp(-2|\vk|/[k_B T (1+\vperp)])$ in the summand, and
Eq.(\ref{lowtemp}).
In Fig.3  we show the energy loss rate per unit volume calculated from the
interpolation formula, divided by the high-temperature rate,
Eq.(\ref{hightemp}).
The conditions assumed are those at the highest density at which nuclei in
the crust are approximately spherical according to the calculations of
Ref.\cite{lrp}, $\mu_e = 78$ MeV and $Z=62$, and form factors are taken to be
unity.  This shows that neutrino emission is suppressed by a factor of 10 or
more for temperatures less than about $10^9$ K.  Including form factors
reduces the total luminosity by more than a factor two, but for $T > 10^9$ K,
the ratio of the interpolation formula result to $\dot{E}_>$
changes by no more than 30\% when form factors are introduced.

We now explore the consequences of the recent discovery that in a
significant fraction of the crustal matter of a neutron star nuclei are likely
to be rod-like (spaghetti) or plate-like (lasagna), and not
spherical\cite{lrp,oya}.  If one neglects spatial variations of the cross
section of the rods or the thickness of the plates, the only reciprocal
lattice vectors that provide scattering in the case of rods will be those that
lie in a plane, while for lasagna the reciprocal lattice vectors must lie on a
line.  To the extent that neutrino emission is dominated by the lowest
reciprocal lattice vectors, this fact implies that neutrino emission for
spherical
nuclei, spaghetti and lasagna would be in the ratio 6:3:1, reflecting the
number of reciprocal lattice vectors for which the form factors do not
vanish.
Other effects
influencing the emission are the dependence on nuclear shape of the magnitudes
of reciprocal lattice vectors, especially the lowest ones, and of form
factors.

Up to now we have assumed that the crystal is perfect, but in reality there may
be impurities and/or lattice imperfections such as dislocations or grain
boundaries.  These will give rise to bremsstrahlung at low temperatures that
will not be suppressed by the band-structure effects that we have considered
above.  Should they be present, their contribution, which varies as $T^6$, will
dominate the loss of energy by neutrino bremsstrahlung at sufficiently low
temperatures.
For the case of impurity scattering in matter with spherical nuclei, the
energy loss by pair bremsstrahlung depends on the impurity concentration,
$x_i$, and the mean square deviation of the atomic number of the impurities
from that of the host lattice, $\langle(\Delta Z)^2\rangle$, and it may be
crudely
estimated to be $\sim x_i \langle(\Delta Z/Z)^2\rangle$ times the
bremsstrahlung rate for
a perfect lattice evaluated neglecting band-structure effects,
Eq.(\ref{hightemp}).  If one uses the estimates of $x_i$ and $\langle(\Delta
Z)^2\rangle$
from Ref.\cite{flowrud}, one finds $x_i \langle(\Delta Z/Z)^2\rangle$ to be
less than
$10^{-5}$, and therefore impurities are unlikely to be important except at
extremely low temperatures, where all other processes vary more rapidly with
temperature than $T^6$.

Absorption of thermally excited phonons on electrons may be accompanied by
neutrino bremsstrahlung.  Flowers\cite{flowers} has estimated the rate of
energy emission by this process for matter with spherical nuclei and finds
that
it varies as $T^{11}$  at temperatures low compared with the Debye
temperature (the ion plasma frequency),  and that at the Debye temperature
it is about 0.1 times the bremsstrahlung rate
estimated neglecting suppression by the band structure effects considered in
this Letter.  At the melting temperature bremsstrahlung from phonons is found
to be comparable to that from the static lattice, a conclusion confirmed by
Itoh et al.\cite{itoh} at densities less that $10^{12}\,{\rm g}\,
{\rm cm}^{-3}$.  We conclude that at low temperatures the phonon process will
dominate bremsstrahlung from the static lattice,
but to make quantitative comparisons it is necessary to recalculate the rate of
the phonon process with allowance for recent developments in the understanding
of matter at sub-nuclear densities, including
the non-spherical nuclear shapes already referred to.

We now assess the importance of neutrino bremsstrahlung from the static lattice
for the thermal
evolution of a neutron star. Consider the case when the result for the
high-temperature limit is applicable.  Our estimates of emissivities for matter
with non-spherical nuclei indicate that these do not exceed those for matter
with spherical nuclei, and therefore we take our estimate of the emissivity of
matter at the highest density at which spherical nuclei exist as an upper
bound on the emissivity of matter in the crust.  The total luminosity of the
crust is thus less than
$\sim 1.1 \times 10^{33} T_8^6 M_{cr}/M_{\odot} \, {\rm erg} \, {\rm s}^{-1}$,
where $M_{cr}$ is the mass of the crust.
According to the calculations of Friman and Maxwell\cite{fm}, the luminosity
due to the
modified Urca process is
$5.3 \times 10^{31} (n_0/n_{core})^{1/3} T_8^8 M/M_{\odot} {\rm erg} \,
{\rm s}^{-1}$ if
neutron superfluidity and proton superconductivity are neglected.
Here $n_{core}$ is the average baryon density in the core,
and $n_0$ is the baryon density of nuclear matter at saturation.
For a typical neutron star with mass $M = 1.4 M_{\odot}$ and radius 10 km,
and for crustal masses taken from Ref.\cite{lrp},
the
modified Urca rates and the neutrino bremsstrahlung rate are comparable only at
temperatures of order $5 \times 10^{7}$ K,
almost one order of magnitude
smaller than indicated by earlier calculations\cite{st}. The effects of band
structure and form factors will minimize the importance of the bremsstrahlung
process still
further, but to determine whether it can ever be important, more detailed
calculations of thermal evolution that allow for nucleon superfluidity and
other effects are required.

To summarize, neutrino-pair bremsstrahlung
from electrons in neutron star crusts is much less important than suggested by
earlier estimates.  One reason for this is that the basic process is
suppressed by band structure effects and a second is
that the amount of matter in the crust of a neutron star is considerably less
than previously estimated.

This work was supported in part by NSF grant NSF PHY91-00283 and NASA grant
NAGW-1583.
We are grateful to D. G. Ravenhall and D. G. Yakovlev for helpful discussions.

\newpage
\section*{Figure Captions}

\noindent
FIG.1. a) The basic bremsstrahlung process.
The double line is the
propagator for a band electron. b) The process in first-order perturbation
theory. The cross denotes an electron-lattice interaction, and the
propagators are free ones.

\medskip \noindent
FIG.2. Electron energy $E$ as a function of $p^{\parallel}$ for
two values of $p^{\perp}$. The arrow shows a possible transition.

\medskip \noindent
FIG.3. Energy emission rate according to the interpolation formula described in
the text compared with the high temperature limit,
$\dot{E}_>$,
Eq.(9),
as a function of temperature.

\end{document}